\documentclass[prl,superscriptaddress,floatfix,showpacs,aps,reprint,letterpaper,nobalancelastpage]{revtex4-1}
\usepackage{graphicx} 
\usepackage{color} 
\usepackage{transparent}
\usepackage{amsmath}
\usepackage{hyperref} 
\usepackage{amssymb}

\newcommand{\ket}[1]{\left\lvert #1 \right\rangle}

\begin{document}

\title{Characterization of addressability by simultaneous randomized benchmarking}\date{\today}

\author{Jay M. Gambetta}
\affiliation{IBM T.J. Watson Research Center, Yorktown Heights, NY 10598, USA}
\author{A. D. C\'orcoles}
\affiliation{IBM T.J. Watson Research Center, Yorktown Heights, NY 10598, USA}
\author{S. T. Merkel}
\affiliation{IBM T.J. Watson Research Center, Yorktown Heights, NY 10598, USA}
\author{B. R. Johnson}
\affiliation{Raytheon BBN Technologies, Cambridge, MA 02138, USA}
\author{John A. Smolin}
\affiliation{IBM T.J. Watson Research Center, Yorktown Heights, NY 10598, USA}
\author{Jerry M. Chow}
\affiliation{IBM T.J. Watson Research Center, Yorktown Heights, NY 10598, USA}
\author{Colm A. Ryan}
\affiliation{Raytheon BBN Technologies, Cambridge, MA 02138, USA}
\author{Chad Rigetti}
\affiliation{IBM T.J. Watson Research Center, Yorktown Heights, NY 10598, USA}
\author{S. Poletto}
\affiliation{IBM T.J. Watson Research Center, Yorktown Heights, NY 10598, USA}
\author{Thomas A. Ohki}
\affiliation{Raytheon BBN Technologies, Cambridge, MA 02138, USA}
\author{Mark B. Ketchen}
\affiliation{IBM T.J. Watson Research Center, Yorktown Heights, NY 10598, USA}
\author{M. Steffen}
\affiliation{IBM T.J. Watson Research Center, Yorktown Heights, NY 10598, USA}
\begin{abstract}
The control and handling of errors arising from cross-talk and unwanted interactions in multi-qubit systems is an important issue in quantum information processing architectures. We introduce a benchmarking protocol that provides information about the amount of addressability present in the system and implement it on coupled superconducting qubits. The protocol consists of randomized benchmarking experiments run both individually and simultaneously on pairs of qubits. A relevant figure of merit for the addressability is then related to the differences in the measured average gate fidelities in the two experiments. We present results from two similar samples with differing cross-talk and unwanted qubit-qubit interactions. The results agree with predictions based on simple models of the classical cross-talk and Stark shifts. 
\end{abstract}
\pacs{03.67.Ac, 42.50.Pq, 85.25.-j}
\maketitle 

The ability to selectively address one qubit or subsystem from among many in a quantum register is a key prerequisite for a scalable quantum computing architecture.  This addressability can be lost either due to control fields that target one subsystem influencing the other neighboring subsystems (classical cross-talk errors), or by unwanted quantum interactions between the target subsystem and the other subsystems. Simple estimates for the classical cross-talk errors can be obtained by measuring the residual Rabi rate on un-targeted qubits \cite{Haffner2005}. Extensions to capture all the relevant error rates are necessary. This issue is particularly important because traditional models for fault-tolerant quantum computation assume uncorrelated errors and therefore these addressability errors need to be characterized and subsequently minimized.

The experimental demonstration of errors of the order required for fault tolerant quantum computation \cite{Shor1996,Aharonov1997,Preskill1997,Knill1998} presents a formidable challenge, as it is typically limited by errors in state preparation and measurement (SPAM).   Indeed,  quantum process tomography (QPT) 
\cite{Chuang1997}, one of the standard methods for characterizing a quantum process, is particularly sensitive to SPAM errors. Additionally, complete quantum-gate characterization rapidly becomes experimentally intractable due to the exponentially large Hilbert space. 

An alternative is randomized benchmarking (RB) \cite{Emerson2005,Knill2008,Magesan2011,Magesan2011a}, which has been used to characterize single qubit gates in liquid-state NMR \cite{Ryan2009}, trapped atomic ions \cite{Knill2008,Biercuk2009,Brown2011,Olmschenk2010}, and superconducting qubits \cite{Chow2009,Chow2010a,Paik2011}. RB is specifically tailored to compensate for SPAM errors by considering only the exponential decay of sequences of random gates. This comes at the cost of only obtaining information about the average gate error over the Clifford 
group, although some alternative approaches have been recently devised for extending RB to estimate the error of a single, particular gate \cite{
Magesan2012,Gaebler2012}.

In this Letter, we present and experimentally implement a method for characterizing the amount of addressability between two subsystems. We typically specialize to qubits, but the method is general and, for example, handles subsystems with logical qubits.  First, we perform RB on each of the subsystems independently, while leaving the other unperturbed. Then we perform (subsystem local) RB on both subsystems simultaneously, and compare 
the error rates.  The comparison of these two can indicate a parametric dependence of the error rate of a subsystem on the operations in another 
subsystem.  Furthermore, we show how the twirling protocol proposed in Ref. \cite{Emerson2007} can be modified with this RB protocol to detect some 
forms of spatial correlations. These methods utilize many of the techniques of Ref. \cite{Magesan2011} and are similarly scalable and less 
dependent on SPAM errors than QPT.

We start by defining a general randomized benchmarking protocol.  We choose a sequence of $m+1$ unitary gates where the first $m$ gates are chosen 
uniformly at random from a group $\mathcal{G} = \{\mathcal{U}_i \}$ and the $(m+1)^{\mathrm{th}}$ gate is chosen to be the inverse of the composition of the 
first $m$ random gates. Provided $\mathcal{G}$ is at most the Clifford group, this undo gate 
can be found efficiently by the Gottesman-Knill theorem \cite{Aaronson2004}. Assuming each gate $\mathcal{U}_{i}$ has some associated error, represented 
by $\Lambda_{i}(\rho)$,  the sequence of gates is modeled by
\begin{equation}
\mathcal{S}_\mathbf{i_m} = \Lambda_{i_{m+1}}\circ\mathcal{U}_{i_{m+1}}\circ \left(\bigcirc_{j=1}^{m}\left[\Lambda_{i_{j}}\circ\mathcal{U}_{i_j}
\right]\right),
\end{equation} 
where $\circ$ represents compositions, $\mathbf{i_m}$ is the $m$-tuple $(i_1,...,i_m)$ describing the sequence, and the recovery gate $\mathcal{U}_{i_{m+1}}$ is uniquely determined by $
\mathbf{i_m}$. Since $\{\mathcal{U}_i\}$ forms a group, the
sequence can be
rewritten as
\begin{equation}
\mathcal{S}_\mathbf{i_m} = \Lambda_{i_{m+1}}\circ \left(\bigcirc_{j=1}^{m}\left[\mathcal{U}^\dagger_{\tilde{i}_j} \circ\Lambda_{i_{j}}\circ\mathcal{
U}_{\tilde{i}_j}\right]\right),
\end{equation}  where $\mathcal{U}_{\tilde{i}_j}$ is another element of $\mathcal{G}$.

For each sequence the overlap $\mathrm{Tr} [ E \mathcal{S}_\mathbf{i_m} (\rho)]$ between $E$, an operator representing the measurement (with errors), and $\mathcal{S}_\mathbf{i_m} (\rho)$, the final state, is measured. Averaging this overlap over $K$ independent sequences of length $m$ gives an estimate of the average sequence fidelity  $F_\mathrm{seq}(m,E,\rho) = \mathrm{Tr} [ E \mathcal{S}_m (\rho)]$ where $\mathcal{S}_m$ is the average sequence 
superoperator given by $\mathcal{S}_{m}=\sum_{\mathbf{i_m}} \mathcal{S}_{\mathbf{i_m}}/K$. Now by defining $\bar\Lambda$ to be the average over $
\Lambda_{{i_j}}$ and by assuming that $\delta \Lambda_{{i_j}} = \Lambda_{{i_j}}-\bar\Lambda$ is small \cite{Magesan2011,Magesan2011a}, the average sequence superoperator can be written as 
\begin{equation}
\mathcal{S}_m  =\bar\Lambda\circ [\mathcal{W}_\mathcal{G}(\bar\Lambda)]^{\circ m} + \mathcal{O}(\delta \Lambda),
\end{equation} 
where $\mathcal{W}$ represents the twirl over the group $\mathcal{G}$ and is given by 
\begin{equation}
\mathcal{W}_\mathcal{G}(\bar\Lambda) = \frac{1}{|\mathcal{G}|}\sum_{\mathcal{U}\in\mathcal{G}} \mathcal{U}^\dagger \circ \bar\Lambda \circ \mathcal{
U}.
\end{equation} Depending on the group $\mathcal{G}$, this map can have a simple structure with a small number of parameters \cite{Merkel2012b}. 

Below we explicitly show the symmetrizing effect of three relevant twirls.  We write a map $\Lambda$ on $n$ qubits using the Liouville representation in the Pauli operator basis, the Pauli transfer matrix (PTM) $\mathcal{R}$, as 
\begin{equation}
(\mathcal{R}_\Lambda)_{ij} =  \frac{1}{d}\mathrm{Tr}[P_i \Lambda (P_j)] 
\end{equation}
where $P_j$ is a Pauli operator for $j\neq0$, $P_0$ is the identity operator $I$, and $d=2^n$ is the dimension of the system \cite{Chow2012}.  This mapping is a matrix representation of quantum maps, defined over the vector space of the Pauli operators, where composition becomes matrix 
multiplication. This allows us to write the twirl as 
\begin{equation}
\mathcal{W}_\mathcal{G}(\mathcal{R}_{\bar\Lambda})^ =\frac{1}{|\mathcal{G}|}\sum_{\mathcal{U}\in\mathcal{G}} \mathcal{R}_{\mathcal{U}}^{\dagger} 
\mathcal{R}_{\bar\Lambda} \mathcal{R}_{\mathcal{U}}.
\end{equation}
The group average over the conjugation action of a group is well known in the representation theory of finite groups and, as a corollary of Shur's 
lemma, would be proportional to the identity if the $\mathcal{R}$ matrices were irreducible (i.e. there is no basis in which each $\mathcal{R}_{
\mathcal{U}}$ is block-diagonal) \cite{Tinkham1964}.  In the case where the representation is reducible the resulting expression contains cross-terms
\cite{Merkel2012b}. In this paper we consider a bipartite system of dimension $d = d_1+d_2$ and twirling over the following groups: $\mathcal{C}$ (the set of Clifford 
operators on the full space), $\mathcal{C}\otimes \mathcal{C}$ (independent Clifford operators on two different subsystems), $
\mathcal{C}\otimes I$ (Clifford operators subsystem 1), and $I\otimes \mathcal{C}$ (Clifford operators on subsystem 
2).

\emph{Full Clifford Twirl} - In the case of the full Clifford group there are two irreducible subspaces corresponding to two stabilized operators: the projector onto $\Pi_0$ (the identity operator) and a projector ($\Pi$) onto the remainder of the Pauli group.  For the full Clifford group
\begin{equation} \label{eq:fulltwirl}
\mathcal{W}_\mathcal{C}( \mathcal{R}_{\bar\Lambda} ) = \begin{pmatrix}1& 0\\
0& \alpha \Pi
\end{pmatrix},
\end{equation}
 which is a depolarizing channel with $\alpha = {{\rm Tr} (\Pi \mathcal{R}_{\bar\Lambda}) }/{{\rm Tr}(\Pi)}$.  If we decompose the initial state and the measurement operator in the Pauli basis, $\rho =\sum_{j} x_j P_i/d$ and $\tilde E = \sum_j\tilde e_jP_j$, (where we have additionally absorbed the error from the final gate into $
\tilde{E}$), then the twirled map sequence fidelity is given by 
\begin{equation}
F_\mathrm{seq}(m,E,\rho) = A\alpha^m +{\tilde e_0},
\end{equation} where $A = \sum_{j\neq 0}{\tilde e_j x_j}$. Provided that $\tilde e_j x_j$  is non-zero (we have an initial state with 
some polarization in the same direction as our measurement), the parameter $\alpha$ can be extracted by fitting the sequence fidelity to a decaying 
exponential. As shown previously \cite{Magesan2011}, the average gate error is $r=(d-1)(1-\alpha)/d$.

\emph{$C_1^{\otimes n}$ Twirl} - When we twirl over the group of subsystem Clifford operators $\mathcal{C}\otimes \mathcal{C}$ we obtain four distinct irreducible subspaces:  $\Pi_0
=I\otimes I$, $\Pi_2=I \otimes \mathbf{P}$, $\Pi_{1}=\mathbf{P} \otimes I$ and $\Pi_{12}=\mathbf{P} \otimes \mathbf{P}$ where $\mathbf{P}$ is the 
vector of Pauli operators for each subsystem. From Shur's lemma we obtain 
\begin{equation}
\mathcal{W}_{\mathcal{C}\otimes \mathcal{C}}( \mathcal{R}_{\bar\Lambda} )=
\begin{pmatrix}
1& 0& 0 & 0\\
0& \alpha_{2|1} \Pi_{2}&0 &0 \\
0& 0& \alpha_{1|2} \Pi_{1}&0 \\
0& 0& 0& \alpha_{12} \Pi_{12}
\end{pmatrix}
\end{equation}
where $\alpha_{k|k'} = {{\rm Tr} (\Pi_k \mathcal{R}_{\bar\Lambda})}/{{\rm Tr} (\Pi_k )}$. This is just a sum over tensor products of depolarizing 
channels, with $\alpha_{k|k'}$ representing the depolarizing effect on subsystem $k$ whilst simultaneously twirling $k'$. We estimate the average gate error for Clifford operations in this context by $r_{k|k'}=(d_{k}-1)(1- \alpha_{k|k'})/d_{k}$ \cite{Merkel2012b}, where $d_k$ is the dimension of subsystem $k$.  If the errors are uncorrelated,  $\bar\Lambda=\bar\Lambda_1\otimes\bar\Lambda_2$, then the correlation coefficient is the product: $\alpha_{12}= \alpha_{1|2}\alpha_{2|1}$.  Consequently, any deviation, $\delta\alpha=\alpha_{12}-\alpha_{1|2} \alpha_{2|1}$, is an indicator of correlations between the subsystems \cite{Emerson2007}. However, some special cases remain undetected by this test \cite{footnote1}. The twirled map sequence fidelity is given by 
\begin{equation}
F_\mathrm{seq}(m,E,\rho) = A_1\alpha_{1|2}^m +A_2\alpha_{2|1}^m + A_{12} \alpha_{12}^m  +{\tilde e_0},
\end{equation} where $A_k = \sum_{j\in \Pi_k}{\tilde e_j x_j}$.  We can easily extract the $\alpha$'s provided we can prepare initial 
states (or measure operators) with support in only one of the irreducible subspaces giving a single exponential decay. Unlike a full Clifford twirl, SPAM errors can have an effect, but are detectable by a deviation from a single exponential.  

\emph{Single Subsystem Twirl} - Twirling over a single subsystem is more complicated.
%as when we only twirl over subsystem $1$ we have $d_2^2$ copies of the single Clifford 
%irreducible subspaces for subsystem $1$. 
The general form is
\begin{equation}
\mathcal{W}_{\mathcal{C}\otimes I}( \mathcal{R}_{\bar\Lambda} )=
\begin{pmatrix}
\mathcal{R}_{\bar\Lambda_2}& 0& 0 \\
0& \Gamma & 0  \\
0& 0& \ddots
\end{pmatrix}\label{eq:subtwril}
\end{equation}
where $\mathcal{R}_{\bar\Lambda_2}$ is the Pauli transfer matrix of the map $\bar\Lambda_2(\rho_2) = \mathrm{Tr}_1[\bar\Lambda(I_1\otimes \rho_2)]/d_1$ and $\Gamma = \sum_{l\neq 0}\left[ \left(P_l\otimes I_2\right) \mathcal{R}_{\bar{\Lambda}} \left(P_l\otimes I_2\right)\right]/(d_1^2-1)$. Due to the block structure of this map, repeated applications 
have the same structure, and tracing over subsystem 2 will give a depolarizing channel on subsystem 1 (a one parameter map 
of the form given in Eq.~\eqref{eq:fulltwirl}). However, the sequence fidelity is not necessarily exponential: after tracing out the second 
subsystem in both measurement and preparation, it is given by 
\begin{equation}
\begin{split}F_\mathrm{seq}(m,E,\rho) =&\tilde e_{0}  + A (\Gamma^m)_{0,0}, 
\end{split}
\end{equation} where $ A = \sum_{j\in \mathbf{P}\otimes I } \tilde e_{j} x_{j}$.  Fortunately, as the error goes to zero, the leading contribution to the $00$ matrix element of $(\Gamma^m)$ is 
$\alpha_1^m$ with  $ \alpha_1 = {{\rm Tr} (\Pi_1 \mathcal{R}_{\bar\Lambda})}/{{\rm Tr} (\Pi_1 )}$. Furthermore, if the data is well approximated by a single exponential decay, then the replacement of $(\Gamma^m)_{0,0}$ by $ \alpha_1^m$ is valid and the 
error on subsystem 1 when nothing is done on subsystem 2 is given by $ r_1=(d_1-1)(1-\alpha_1)/d_1$. We can obtain a similar estimate for $r_2$ with the twirl $\mathcal{W}_{\mathcal{C}\otimes I}$. 

We can then quantify the increase in error rate from simultaneous control. This gives us a metric of addressability: the additional errors induced on subsystem $k$ from controlling $k'$,
\begin{equation}
\delta r_{k|k'}= \left| r_k - r_{k|k'} \right|.
\label{addressabilityMetric}
\end{equation}  This metric combines the errors from classical cross-talk leakage errors and quantum coupling between the systems.

\begin{figure}[t!]
\centering
\includegraphics[width=0.48\textwidth]{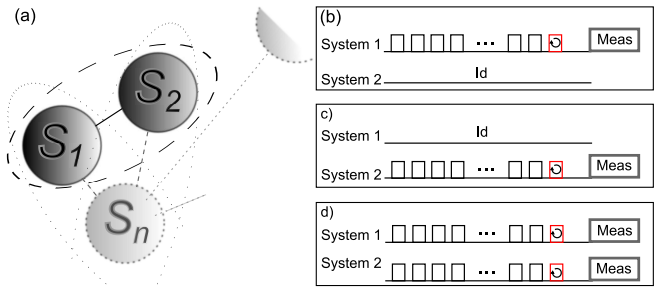}
\caption{\label{fig:1} (color online). (a) General approach to characterize the effect of cross-talk and unwanted quantum interactions in a processor with multiple subsystems. The experiments are performed pairwise. For each pair, three experiments are performed: RB is performed on each subsystem while leaving the other subsystem unperturbed (b and c); then RB is performed simultaneously (d).}
\end{figure}

In short, the addressability protocol is the following three experiments: 

\underline{Experiment 1}: Implement RB on the first subsystem, i.e. twirling with the group $\mathcal{C}\otimes I$ (see Fig.~\ref{fig:1}b). Fit the average decay of subsystem 1's initial state to obtain $\alpha_1$ and thus the error $r_1$.

\underline{Experiment 2}: Perform the same experiment on subsystem 2 (see Fig.~\ref{fig:1}c) yielding $\alpha_2$ and $r_2$. 

\underline{Experiment 3}: Implement RB on both subsystems simultaneously, i.e. twirl with $\mathcal{C}\otimes \mathcal{C}$, (see Fig.~\ref{fig:1}d).  Fit the decays of the single subsystems (e.g. $ZI$ and $IZ$) and two-qubit correlations (e.g. $ZZ$) to obtain $\alpha_{1|2}$,  $\alpha_{2|1}$ and $\alpha_{12}$.  The addressability is then quantified from Eq. \ref{addressabilityMetric}, and potential correlations in the errors are flagged by examining $\delta\alpha=\alpha_{12}-\alpha_{1|2}\alpha_{2|1}$.

To demonstrate this protocol we implement it in two samples, $a$ and $b$, with identical layouts. The samples consist of two single-junction transmons (SJT) \cite{Chow2012} coupled by a coplanar waveguide resonator.  We perform single-qubit rotations by applying microwave pulses along individual qubit drive lines resonant with the qubit frequency. The pulses were shaped for high-fidelity control \cite{Motzoi2009}. The pulses applied to each qubit belong to the single-qubit Clifford group, generated by $\left\{X_{\pm\pi/2}, Y_{\pm\pi/2}, X_{\pi}, Y_{\pi}\right\}$ where $R_{\theta}$ represents a rotation of angle $\theta$ around $R$. 

The design of all SJTs are as described in Ref.~\cite{Chow2012}. The SJTs on sample $a(b)$ have transition frequencies $\omega_{1}/2\pi=4.9895$ (4.7610) and $\omega_{2}/2\pi=5.0554$ (5.3401) GHz, with relaxation times of $T_1^1=9.7$ (9.4) and $T_1^2=8.2$ (9.9) $\mu$s and coherence times of $T^{*1}_2=10.3$ (7.3) and $T^{*2}_2=7.1$ (10.2) $\mu$s. The cavity of sample $a(b)$ resonates at 7.325 (7.4269) GHz. The samples are radiation shielded \cite{Corcoles2011} and thermally anchored to 15 mK of a dilution refrigerator. Sample $a$ is the same device as used in Ref.~\cite{Chow2012} where QPT performed on various single qubit gates revealed an average gate error of $\sim 3.6\%$.

\begin{figure}[t!]
\centering
\includegraphics[width=0.48\textwidth]{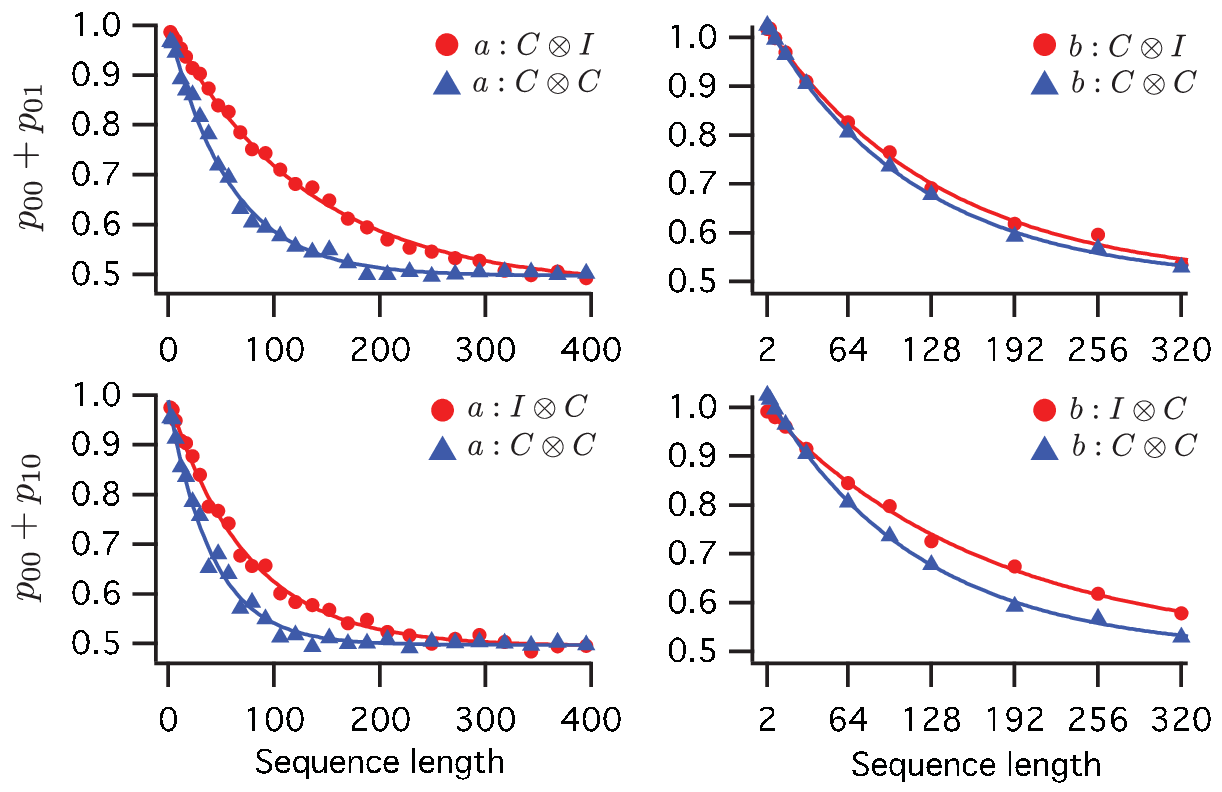}
\caption{\label{fig:2} (color online) Experimental results of RB experiments on sample $a$ (left column) sample $b$ (right column). Circles (red) are individual RB fidelity decays, triangles (blue) are simultaneous RB decays.  Top row shows an effective projection into qubit 1's subspace, bottom row, into qubit 2.  Fits to single exponential decays (solid lines) give a maximum reduced $\chi^2 = 1.63$, demonstrating  a reasonable fit model. The significant increase in decay rate for sample $a$ under simultaneous RB witnesses addressability errors.}
%\clearpage
\end{figure}

Each RB experiment starts with the qubits in the ground state $\ket{00}$.  In order to simulate tracing over the other qubit, a set of experiments is run with additional rotations to measure the final populations $p_{00}$, $p_{01}$, $p_{10}$, $p_{11}$.  From the decay of $p_{00} + p_{01}$ and $p_{00} + p_{10}$ we simulate tracing over qubit 2 and 1, respectively. The results of the three experiments are shown in Fig.~\ref{fig:2} and in Table~\ref{table:1}.  We find a single exponential decay is a good fit to the data, with a maximum reduced $\chi^2$ of 1.63 (most fits have $\chi^2 < 1$), and thus the addressability RB protocol is valid (for more information on the fitting procedure see the supplementary material \cite{Merkel2012b}). From these experiments the addressability error in sample $a$ is $\sim0.5\%$ whereas in sample $b$ the error is indistinguishable from zero. Furthermore, from the decay of $p_{00}+p_{11}$, we can obtain $\alpha_{12}$ (Table~\ref{table:1}) and for sample $a$ this is substantially different from $\alpha_{1|2}\alpha_{2|1}$ which is consistent with correlated errors.

\begin{table}
\begin{ruledtabular}
\begin{tabular}{cc|cc}
& Twirl Group & Sample $a$ & Sample $b$ \\
\hline
$r_1$ & $C\otimes I$ & $0.0039\pm 0.0001$ & $0.0029 \pm 0.0002$\\
$r_2$ & $I\otimes C$ & $0.0067 \pm 0.0002 $ & $0.0037 \pm 0.0003$ \\
$r_{1|2}$ & $C\otimes C$ & $0.0086 \pm 0.0003$& $0.0032 \pm 0.0003$ \\
$r_{2|1}$ & $C\otimes C$ & $0.0120\pm 0.0005$ & $0.0043 \pm 0.0002$ \\
$\delta r_{1|2}$ & - & $0.0047 \pm 0.0003 $ & $0.0003 \pm 0.0003$ \\
$\delta r_{2|1}$ & - & $0.0053 \pm 0.0005 $ & $0.0006 \pm 0.0003$ \\
$\delta \alpha$ & - & $0.0050 \pm 0.0018 $ &  $0.0015 \pm 0.0007 $\\
\end{tabular}
\end{ruledtabular}
 \caption{\label{table:1} Summary of the extracted data from the three RB experiments on the two samples in terms of effective error rates and addressability metrics.  Uncertainties are reported as $1\sigma$ confidence intervals. }
\end{table}

We expect the qubit frequency spacing to play a large role in addressability. We do not have perfect classical addressing of the qubits; therefore, we expect the off-resonant drive of one qubit's control pulse to give an AC Stark shift on the untargeted qubit.  In addition, the strength of the cavity-mediated coupling is inversely proportional to the qubit-qubit detuning.   In our experiment we expect these effects to be much stronger in sample $a$ because the qubit-qubit detuning is $\Delta/2\pi= -66$ MHz,  whereas in sample $b$,  $\Delta/2\pi= -579$ MHz. This difference is successfully captured by our metrics.

In our system the qubits are coupled by an $XX$ interaction, with strength $J$, so the Hamiltonian  is written in the dressed basis as  
\begin{equation}
\begin{split}
H =&\varepsilon_1 \left( XI +(m_{12}-\nu_1) IX  - \mu_1 ZX + m_{12} \mu_2 XZ \right)   \\
&+\varepsilon_2 \left( IX +(m_{21}+\nu_2) XI  + \mu_2 XZ -m_{21} \mu_1 ZX \right) \\
&- \omega_1 ZI/2  -\omega_2 IZ/2 +\zeta ZZ/4 , 
\end{split}
\end{equation} where  $\varepsilon_{1(2)}$ are the shaped microwave amplitude of a drive applied on qubit 1(2), $m_{12(21)}$ represent spurious crosstalk due to stray electromagnetic coupling in the device circuit and package, $\mu_{1(2)}$ are the cross resonance coupling parameters ($\mu_{1(2)}=J/\Delta$ for ideal qubits \cite{Rigetti2010}), and  $\zeta$ is an energy shift which can occur due to couplings in the higher levels of the SJTs \cite{DiCarlo2009}. These parameters can be measured by a series of $\pi$-Rabi and $\pi$-Ramsey experiments (standard single qubit Rabi and Ramsey experiments conditioned on the state of the other qubit) 
and we find for sample $a$, 
$m_{1} = 0.19$, $m_{2}=0.32$, $\mu_1=-0.088$, $\mu_2=-0.16$, $\nu_1=-0.025$, $\nu_2=-0.048$, and $\zeta/2\pi=1.1$ MHz. From these values we estimate $\delta r_{1|2}=0.0034$ and $\delta r_{2|1}= 0.007$. These estimates are of the same order as the measured values, though it is likely that this simple model does not capture all addressability errors. We note however, that despite the higher addressability error for sample $a$, a high two-qubit gate fidelity ($95\%-98.5\%$) was measured~\cite{Chow2012}, suggesting that the tradeoffs between gate fidelities and addressability will need to be further explored. 

In conclusion, we have presented a protocol to measure the addressability of the gates used in a quantum information processing device. We apply this protocol
to two samples of two SJT coupled by a co-planar waveguide resonator.  In the first sample the single qubit average gate error is measured to be $\sim 0.5\%$ and with
operations on the second qubit this increases to $\sim 1.0\%$ (an addressability error of $\sim 0.5\%$) indicating that benchmarking a single component does not necessarily account for its effect on the 
larger device. Furthermore, QPT on this device gave an even larger average single qubit gate error of $\sim3.6\%$ which we attribute to the sensitivity of QPT to SPAM errors.  The second device had an equivalent average gate error, but much smaller addressability error due to a larger detuning between the qubits.

\begin{acknowledgments}
We acknowledge discussions and contributions from Easwar Magesan, George A. Keefe, Mary B. Rothwell, J. R. Rozen, Marcus P. da Silva and Joseph Emerson.
We acknowledge support from IARPA under contract W911NF-10-1-0324. All statements of fact, opinion or conclusions contained herein are those of the 
authors and should not be construed as representing the official views or policies of the U.S. Government.
\end{acknowledgments}

%\bibliographystyle{JMGbibstylePRL}
%\bibliography{JMGBooks,JMGArticles,Arxivpapers}

\begin{thebibliography}{10}
\bibitem{footnote1}
 It is easy to come up with contrived examples where correlated errors give $\alpha_{12}=\alpha_{1|2} \alpha_{2|1}$, e.g. the error is a CNOT gate.
\end{thebibliography}

\end{document}

\title{Supplementary material for `Characterization of addressability by simultaneous randomized benchmarking'}
\date{\today}

\author{Jay M. Gambetta}
\affiliation{IBM T.J. Watson Research Center, Yorktown Heights, NY 10598, USA}
\author{A. D. C\'orcoles}
\affiliation{IBM T.J. Watson Research Center, Yorktown Heights, NY 10598, USA}
\author{S. T. Merkel}
\affiliation{IBM T.J. Watson Research Center, Yorktown Heights, NY 10598, USA}
\author{B. R. Johnson}
\affiliation{Raytheon BBN Technologies, Cambridge, MA 02138, USA}
\author{John A. Smolin}
\affiliation{IBM T.J. Watson Research Center, Yorktown Heights, NY 10598, USA}
\author{Jerry M. Chow}
\affiliation{IBM T.J. Watson Research Center, Yorktown Heights, NY 10598, USA}
\author{Colm A. Ryan}
\affiliation{Raytheon BBN Technologies, Cambridge, MA 02138, USA}
\author{Chad Rigetti}
\affiliation{IBM T.J. Watson Research Center, Yorktown Heights, NY 10598, USA}
\author{S. Poletto}
\affiliation{IBM T.J. Watson Research Center, Yorktown Heights, NY 10598, USA}
\author{Thomas A. Ohki}
\affiliation{Raytheon BBN Technologies, Cambridge, MA 02138, USA}
\author{Mark B. Ketchen}
\affiliation{IBM T.J. Watson Research Center, Yorktown Heights, NY 10598, USA}
\author{M. Steffen}
\affiliation{IBM T.J. Watson Research Center, Yorktown Heights, NY 10598, USA}
\maketitle

\section{Parameter estimation: a statistical analysis}

The experimental data is fit to a three parameter model using standard 
non-linear least squares regression. We use the  \texttt{nlinfit} routine provided by Matlab \cite{Matlab}. Uncertainties are identified from 
estimates of the Jacobian at the best-fit point with a linear 
assumption. To do this we use the Matlab routine \texttt{nlparci} with the confidence set to $68\%$ ($1\sigma$) \cite{Matlab}.   Identifying goodness of fit is somewhat muddled because of 
open questions about the underlying distribution of sequence fidelities. 
There is no guarantee that random sequence fidelities follow a normal 
distribution.  Indeed, anecdotally we observe clustering of high and low 
fidelities in some data sets. Furthermore, for very high sequence 
fidelities we see skewed distributions because the fidelity is bounded by 
one. Nevertheless, the residuals from fits of the average fidelities 
appear to follow a normal distribution and as such we can use a chi-square test for the \textit{goodness of fit}. The reduced chi-square  is given by $\chi^2/\mathrm{dof}$, where $\mathrm{dof}$ is the number of degrees of freedom and $\chi^2$ is the chi-square given by 
\begin{equation}
\chi^2 = \sum_{i=1}^N \frac{ (x_i-\mu_i)^2}{\sigma_i^2}.
\end{equation} Here $N$ is the number of  experimentally measured quantities $x_i$,  $\mu_i$ is the expected value from the model under test and $\sigma_i^2$ is the expected variance in the measured quantities. In this case we take $N$ to be the number of truncations used, $x_i$ is the experimentally obtained sequence fidelity after averaging over the different random Clifford strings, and we use the standard error in this estimate of the sequence fidelity as $\sigma_i$.

For sample $a$ there are 32 different truncations and with the 3 fit parameters (29 degrees of freedom) we find for $(\alpha_1, \alpha_2, \alpha_{1|2}, \alpha_{2|1}, \alpha_{12})$ the reduced  chi-square are (1.567, 0.809, 1.369, 1.639, 1.189) respectively. The data for sample $b$ have 11 different truncations which gives 8 degrees of freedom and we find the reduced chi-square for these fits are (0.638, 1.609, 0.156, 0.489, 0.280) respectively. All these reduced chi-square's are close to or less than one so we conclude that the model is a good fit to the data.

\section{Review of group theory}
In this section we review some of the standard results of the representation theory of finite groups and use this to decompose the PTM's of the various groups from the main text into their irreducible components.  This decomposition greatly facilitates twirling (i.e.~the group average). 

\subsection{Representation theory of finite groups}   
  
Since we will extensively be using Schur's lemma later in this document we will give a very brief overview of representation theory for finite groups.  A group is a set $\mathcal{G}$ and an operation $\circ$ with the following properties: 
\begin{itemize} 
\item{\bf closure} for all $g_1, g_2 \in \mathcal{G}$, $g_3 = g_1 \circ g_2 $ is also an element of $\mathcal{G}$,
\item {\bf associativity} for all $g_1, g_2, g_3 \in \mathcal{G}$  $(g1 \circ g_2) \circ g_3 = g_1 \circ( g_2 \circ g_3)$,
\item {\bf identity} there exists an element $I \in \mathcal{G}$ such that for all $g \in \mathcal{G}$, $I \circ g = g\circ I = g$,
\item {\bf inverse} for all $g \in \mathcal{G}$ there exists a $g^{-1} \in \mathcal{G}$ such that $g g^{-1} = g^{-1} g = I$.
\end{itemize}
In the remainder of this document we will take the group operation, $\circ$, to be multiplication unless otherwise stated.  

A representation of a group is a map $\sigma$ from group elements to a set of complex matrices, $\sigma : \mathcal{G} \rightarrow \mathbb{C}^{d \times d}$, such that the group operation is preserved under matrix multiplication of the representation, i.e.~$\sigma(g_1) \sigma(g_2) = \sigma(g_1 g_2)$.  Note that multiplication can be preserved even if $\sigma$ is not a one to one mapping.  For example, all groups have a trivial representation where $\sigma(g) = 1$ since $\sigma(g_1) \sigma(g_2) = \sigma(g_1 g_2)$ is always satisfied by $1 \times 1  = 1$.  A representation that maps each group element to a unique matrix is known as a faithful representation.  Additionally, any matrix representation of a group with non-vanishing determinants is equivalent under a similarity transformation, $\sigma(g) \rightarrow S \sigma(g) S^{-1}$, to a unitary representation, $\sigma(g)^{-1} = \sigma(g)^\dagger $ \cite{Tinkham}.

The final requirement before stating Schur's lemma is to define the notions of reducible and irreducible representations.  A representation is reducible if there exists a similarity transform $S$ such that each group element is block diagonal.  That is, for all $g \in \mathcal{G}$,
\begin{equation}
S \sigma(g) S^{-1} = \left(\begin{array}{cc}
\sigma_1(g) & \mathbf{0} \\
\mathbf{0} & \sigma_2(g)
\end{array} \right).
\end{equation}
If there is no such $S$ then the representation is irreducible.  We traditionally define two representations as distinct if there is no similarity transformation that maps one to the other.  Any general reducible representation can be written as a direct sum of its irreducible components.  Mathematically, representations are more commonly defined by the vector space the $\sigma$'s are acting on, and so we will generally talk of a reducible representation being composed of irreducible subspaces which are preserved, or stabilized, by multiplication by elements of the representation.

There are many ways to state Schur's lemma, but a particularly transparent form for the follwoing is from Tinkham (which also contains detailed proofs) \cite{Tinkham}.  It is split into two components.  

{\lemma Schur's Lemma: The only matrices that commute with all elements of an irreducible representation of a group  are constant matrices (i.e.~scalar multiples of I).\label{L:Schur}}
{\lemma  Given two irreducible representations $\sigma_1$ and $\sigma_2$ and a rectangular matrix $M$ that satisfies $M \sigma_1(g) = \sigma_2(g) M$, for all $g \in \mathcal{G}$, either $M = \mathbf{0}$ or $M$ is invertible and thus $\sigma_1$ and $\sigma_2$ are equivalent modulo a similarity transformation. \label{L:Schur_extend}}

We can now discuss the implications Schur's lemma has to twirling in three increasingly more general corollaries.
{\corollary If $\sigma$ is an $d$-dimensional irreducible representation of $\mathcal{G}$ then for any matrix $M \in \mathbb{C}^{d \times d}$ \label{C:irred}}
\begin{equation}
\frac{1}{|\mathcal{G}|}\sum_{g \in \mathcal{G}} \sigma(g) M \sigma(g)^{-1} = \frac{\textrm{Tr}(M)}{d} I \label{eq:corollary}
\end{equation}
 \begin{proof} The sum on the left hand side of Eq.~\eqref{eq:corollary} commutes with $\sigma(\tilde{g})$ for $\tilde{g} \in \mathcal{G}$.  Explicitly,
 \begin{equation}
 \begin{split}
 \sigma(\tilde{g}) \left( \sum_{g \in \mathcal{G}} \sigma(g) M \sigma(g)^{-1}\right) =&  \sum_{g \in \mathcal{G}} \sigma(\tilde{g} g) M \sigma\left(g^{-1}\right),\\
  =&  \sum_{g' \in \mathcal{G}} \sigma(g') M \sigma\left(g'^{-1} \tilde{g} \right),\\
  =&  \sum_{g' \in \mathcal{G}} \sigma(g') M \sigma\left(g'^{-1}  \right) \sigma(\tilde{g}),\\
  =& \left( \sum_{g \in \mathcal{G}} \sigma(g) M \sigma\left(g  \right)^{-1}\right) \sigma(\tilde{g}),\\
 \end{split}
 \end{equation}   
 where in the intermediate steps we have taken $g' = \tilde{g} g$. Due to Schur's lemma we know that any operator that commutes with all elements of the irreducible representation of a group must have the form
 \begin{equation}
 \frac{1}{|\mathcal{G}|}\sum_{g \in \mathcal{G}} \sigma(g) M \sigma(g)^{-1} = \alpha I,
 \end{equation}
 and to solve for $\alpha$ we take the trace of both sides yielding
 \begin{equation}
 \textrm{ Tr}(M) = \alpha d.
 \end{equation}
\end{proof}

A generalization of the previous corollary is 

{\corollary   If $\sigma$ is a direct sum of distinct irreducible representations of $\mathcal{G}$, $\sigma_1 \oplus \sigma_2 \oplus \ldots \sigma_m$ then for any matrix $M \in \mathbb{C}^{d \times d}$
\begin{equation}
\frac{1}{|\mathcal{G}|}\sum_{g \in \mathcal{G}} \sigma(g) M \sigma(g)^{-1} = \sum_{j=1}^m \frac{{\rm Tr} (M \mathbb{P}_j)}{{\rm Tr}(\mathbb{P}_j)} \mathbb{P}_j, \label{eq:corollary2}
\end{equation}
where $\mathbb{P}_j$ is the projector onto the irreducible subspace defined by $\sigma_j$.\label{C:reduc}}
\begin{proof} 
As in the previous case, the sum on the left hand side of Eq.~\eqref{eq:corollary2} (which for compactness of notation we will label $\Phi = \sum_{g \in \mathcal{G}} \sigma(g) M \sigma(g)^{-1}/|\mathcal{G}| $) commutes with $\sigma(\tilde{g})$ for all $\tilde{g} \in \mathcal{G}$.  Furthermore,  
\begin{equation}
\begin{split}
\mathbb{P}_i \sigma(\tilde{g})\Phi \mathbb{P}_j=&\mathbb{P}_i \Phi \sigma(\tilde{g}) \mathbb{P}_j\\
 \sigma_i(\tilde{g})\mathbb{P}_i \Phi \mathbb{P}_j=&\mathbb{P}_i \Phi \mathbb{P}_j \sigma_j(\tilde{g}) 
\end{split}
\end{equation} 

The matrix $\mathbb{P}_i \Phi \mathbb{P}_j$ is now subject to the constraints of Lemma \ref{L:Schur_extend} and thus 
\begin{equation}
\mathbb{P}_i \Phi \mathbb{P}_j = \delta_{i,j} \alpha_i  \mathbb{P}_i,
\end{equation}
and as in Corollary \ref{C:irred} we can take the trace of both sides to obtain
\begin{equation}
 \textrm{ Tr}(M\mathbb{P}_i) = \alpha_i \textrm{ Tr}(\mathbb{P}_i).
\end{equation}

\end{proof}

Finally, we can extend this corollary to the case of a completely general reducible representation.  Up to a similarity transformation a general representation can be expressed as a direct sum of $n_j$ copies of each of the irreducible representations $\sigma_j$.  We will explicitly write the subspace spanned by the $k^{\rm th}$ copy of representation $\sigma_j$ as being spanned by the set of vectors $\ket{v_{j,k,l}}$.  With this convention  
{\corollary Given a general representation $\sigma$ of $\mathcal{G}$ then for any matrix $M \in \mathbb{C}^{d \times d}$

\begin{equation}
\begin{split}
\frac{1}{|\mathcal{G}|}\sum_{g \in \mathcal{G}} \sigma(g) &M \sigma(g)^{-1} =  \\
&\sum_{j} \sum_{k,k' = 1}^{n_j} \frac{{\rm Tr}(\mathbb{Q}_{j,k,k'}^T  M )}{{\rm Tr}(\mathbb{Q}_{j,k,k'}^T\mathbb{Q}_{j,k,k'}   )} \mathbb{Q}_{j,k,k'}. \label{eq:corollary3}
\end{split}
\end{equation}
where $\mathbb{Q}_{j,k,k'}=\sum_{l } \ket{v_{j,k,l}} \bra{v_{j,k',l}}$
\label{C:general}}
\begin{proof} 

This proof proceeds along the lines of the last but now we use the projectors $\mathbb{P}_{j,k} = \sum_{l } \ket{v_{j,k,l}} \bra{v_{j,k,l}}$ and find that
\begin{equation}
\mathbb{P}_{j,k} \Phi \mathbb{P}_{j',k'} = \delta_{j,j'} \alpha_{j,k,k'}  \mathbb{Q}_{j,k,k'},
\end{equation}
If we multiply both sides by $\mathbb{Q}_{j,k,k'}^T$ and trace we find 
\begin{equation}
{\rm Tr}(\mathbb{Q}_{j,k,k'}^T  M )= \alpha_{j,k,k'}  \textrm{ Tr}(\mathbb{Q}_{j,k,k'}^T\mathbb{Q}_{j,k,k'}).
\end{equation}
\end{proof}

The last ingredient from the representation theory of finite groups that we will require is the following.  If we have a group $\mathcal{G}$ with irreducible representations $\{\sigma_1, \sigma_2, \ldots \}$, we can form the irreducible representations of the group $\mathcal{G} \times \mathcal{G}$ (with elements $g_ig_1, g_1g_2, g_2g_1, \ldots $)by the tensor products $\{\sigma_1\otimes\sigma_1, \sigma_1\otimes\sigma_2,\sigma_2\otimes\sigma_1, \ldots \}$ \cite{Tinkham}.  This is of particular interest when we consider composite Hilbert spaces with many subsystems. 
\subsection{Decomposition of some useful unitary groups}

\subsubsection{Pauli group}
Let us consider the set of channels $\mathcal{P}$ with elements of the form $\Lambda_k(\rho) = P_k \rho P_k$ and derive a general form for the PTM.  Here, as in the text, $P_k$ are the n qubit Pauli operators $\{I,X,Y,Z \}^{\otimes n}$ with $I = \left(\begin{array}{cc} 1&0\\0&1 \end{array} \right)$,$X = \left(\begin{array}{cc} 0&1\\1&0 \end{array} \right)$,$Y= \left(\begin{array}{cc} 0&-i\\i&0 \end{array} \right)$ and $Z = \left(\begin{array}{cc} 1&0\\0&-1 \end{array} \right)$.  As a convention we will take $P_0 = I^{\otimes n}$.

   First, Pauli operators either commute or anticommute, so $P_k P_j P_k = \pm P_j$ and thus 
\begin{equation}
(\mathcal{R}_{\Lambda_k})_{ij} = \pm \frac{ 1}{2^n}\textrm{Tr} \left(P_i P_j  \right) = \pm \delta_{ij}.
\end{equation}
Furthermore, by expressing mapping the four Pauli operators to a pair of bits $(vw)$ according to $\{I,X,Y,Z\} \rightarrow \{00,01,10,11\}$ (and therefore $P_j$ to two binary vectors $(\mathbf{v}, \mathbf{w})$) we find that for the channel above the elements of the PMT are 
\begin{equation}
(\mathcal{R}_{\Lambda_k})_{ii} = (-1)^{\mathbf{v}^{(i)}\cdot\mathbf{w}^{(k)}+\mathbf{w}^{(i)}\cdot\mathbf{v}^{(k)} }.
\end{equation}
All the PMT's are diagonal with half the elements $1$ and half $-1$, except for the identity which is mapped to all $1$'s.  

This is somewhat counter-intuitive, since we have represented the Pauli operators as a set of commuting matrices.  Additionally, while the $4^n$ matrices $P_j$ do not themselves form a group under matrix multiplication the $\mathcal{R}_{\Lambda_k}$ matrices do.  The essential ingredient is that the set of channels $\mathcal{P}$ forms a group under composition because the conjugation action of a unitary channel makes them insensitive to a global phase (i.e.~it is irrelevant whether $X$ times $Y$ is equal to $\pm iZ$ for channels).

We can now explore the structure of the $\mathcal{R}_{\Lambda_k}$ representation of $\mathcal{P}$ in a little more detail.  Group multiplication $\mathcal{R}_{\Lambda_k} =\mathcal{R}_{\Lambda_i}\mathcal{R}_{\Lambda_j}$ is described by $(\mathbf{v}^{(k)}, \mathbf{w}^{(k)}) = (\mathbf{v}^{(j)} \oplus \mathbf{v}^{(i)},\mathbf{w}^{(j)} \oplus \mathbf{w}^{(i)})$, and is simply a representation of $\mathbb{Z}_2^{~\otimes 2n}$.  In fact, the $\mathcal{R}_{\Lambda_k}$ matrices are what is known as the regular representation of $\mathbb{Z}_2^{~\otimes 2n}$ which is defined as the representation over the vector space of the group elements themselves.  For $\mathbb{Z}_2^{~\otimes 2n}$ the regular representation is the direct sum of all $2^{2n}$ distinct irreducible representations.

\subsubsection{Clifford group}
The Clifford group $\mathcal{C}$ is the normalizer of the Pauli group in the space of unitary channels, that is, conjugation by an element of the Clifford group at most permutes the elements of the Pauli group or $C_j P_i C_j^\dagger = \pm P_k$ for all Pauli operators $P_i$.  Therefore, the PTM's describing elements of the Clifford group have a $1$ or $-1$ in each row and column  with the remaining elements zero.  Furthermore, conjugation preserves commutator relations, or alternatively quantum channels are completely positive, which places additional restrictions on the set of physical ${\cal R}$ matrices.  

We will now show that the PTM representation of the Clifford group is of the form $\mathbb{I} \oplus \sigma$, i.e. a direct sum of the trivial representation of the Clifford group and an irreducible representation we will label $\sigma$. The trivial irreducible subspace corresponds to the identity ($P_0 = I$) which is left invariant by all Clifford channels.  To show that the remaining subspace is irreducible we note the the $\mathcal{P}$ is a subgroup of $\mathcal{C}$.  The implication from the previous section is that the only matrices that commute with all elements of the representation are diagonal.  The only restriction on how the Clifford group permutes the Pauli group is that it preserves commutator relations so for any two Pauli matrices $P_i$ and $P_j$ ($i,j \neq 0$)there exists an element of the Clifford group, $C$, such that $C P_i C^\dagger = P_j$.  Therefore no subspace of the Pauli's is stabilized except for the entire group (minus the identity).

For a composition of Clifford operators acting indenpendently on $n$-subsystems, $\mathcal{C}^{\otimes n}$, the irreducible representations are of the form $(\mathbb{I} + \sigma)^n$.  As an example, the group $\mathcal{C}\otimes \mathcal{C}$ has the following irreducible representations: $\mathbb{I}\mathbb{I} = \{II\}$, $\mathbb{I}\sigma = \{IX,IY,IZ\}$, $\sigma\mathbb{I} = \{XI,YI,ZI\}$ and $\sigma\sigma = \{XX,XY,XZ,YX,YY,YZ,ZX,ZY,ZZ\}$.  Addionally we could consider the Cliford group acting on a subsystem of a large space.  In this case we are still representing the single subsystem Clifford group, but with degeneracy.  For example, $I\otimes \mathcal{C}$ has four copies of the irreducible representation $\mathbb{I}$  ($\{II\}$,$\{XI\}$,$\{YI\}$ and $\{ZI\}$) and four copies of $\sigma$ ($\{IX,IY,IZ\}$,$\{XX,XY,XZ\}$,$\{YX,YY,YZ\}$ and $\{ZX,ZY,ZZ\}$).

\subsubsection{Unitary group}
The structure of the full unitary group is essentially identical to that of the Clifford group, that is the PTM representation of a general unitary channel has the form $\mathbb{I} \oplus \sigma$, where the identity is stabilized and the remaining subspace is irreducible.  This should come as no surprise since the PTM mapping is quadratic in the unitary matrices and the Clifford group is a 2-design of the unitary group.  In fact, requiring that all but the identity forms an irreducible subspace is equivalent to the statement that the Clifford group is a 2-design \cite{Gross}.

\subsection{Twirling}

Twirling a channel $\Lambda$ over a group of unitary channels $\mathcal{G} = \{ \mathcal{U}_1,\mathcal{ U}_2, \ldots  \}$ is described by
\begin{equation}
\mathcal{W}_{\mathcal{G}}(\Lambda) = \frac{1}{|\mathcal{G}|}\sum_{\mathcal{U}_j \in \mathcal{G}} \mathcal{U}_j^\dagger \circ \Lambda \circ \mathcal{U}_j ,
\end{equation}
which in the $\mathcal{R}$ representation is 
\begin{equation}
\mathcal{W}_{\mathcal{G}}(\mathcal{R}_\Lambda) = \frac{1}{|\mathcal{G}|}\sum_{\mathcal{U}_j \in \mathcal{G}} \mathcal{R}_{\mathcal{U}_j}^T\mathcal{R}_\Lambda\mathcal{R}_{\mathcal{U}_j} 
\end{equation}
With a detailed understanding of the representation structure of the Pauli transfer matrices the results of twirling are an almost trivial consequence of Corollaries \ref{C:irred} \ref{C:reduc} and \ref{C:general}.

\subsubsection{Pauli twirling}

The PTM representation of the Pauli group is a direct sum of $4^n$ 1-dimensional distinct irreducible representations (i.e.~ the Pauli operators themselves are each stabilized by each element of the Pauli group in this representation since the group is Abelian).  Therefore twirling over the Pauli group is a direct consequence of Corollary 2 and is described    
\begin{equation}
\mathcal{W}_{\mathcal{P}} (R_{\Lambda})_{ij} = \delta_{ij} (R_{\Lambda})_{ii}.
\end{equation}  
Only the diagonal elements of $\mathcal{R}$ remain. This is exactly the space of Pauli channels, that is channels of the form $\sum_j p_j \Pi_j$.

\subsubsection{Clifford twirling}

The Clifford group has  a representation that is the sum of two irreducible blocks, the identity and everything else. For any trace-preserving map the element of $\mathcal{R}$ mapping the identity to itself is unity and therefore we can use Corollary 2 again to find
\begin{equation}
\mathcal{W}_{\mathcal{C}} (R_{\Lambda})_{ij}= \delta_{ij} \left\{\begin{array}{ccc}
1 &{\rm if}~ i=0 \\
({\rm Tr} ({\cal R}_{\Lambda}) - 1)/(d^2-1)& {\rm otherwise}. \\
\end{array} \right.
\end{equation}
This is the set of depolarizing channels, $\Lambda_{\rm dep} (\rho) = \alpha \rho + (1-\alpha)I$ where the survival probability $\alpha = \frac{{\rm Tr} ({\cal R}_\Lambda) - 1}{4^n-1}$.

\subsubsection{Subsytem Clifford twirling}

Here we will focus on a two qubit system and look at twirling over the groups $\mathcal{C} \otimes \mathcal{C}$, $\mathcal{C} \otimes I$ and $I \otimes \mathcal{C}$.  The simplest is $\mathcal{C} \otimes \mathcal{C}$ which we previously determined has a PTM that is a direct sum of four disntinct irreducible representations: $\mathbb{I}\mathbb{I}$,$\mathbb{I} \sigma$, $\sigma\mathbb{I}$ and $\sigma \sigma$.  If we define $\mathbb{P}_{\mathbb{I}\mathbb{I}}$, $\mathbb{P}_{\mathbb{I}\sigma}$, $\mathbb{P}_{\sigma\mathbb{I}}$ and $\mathbb{P}_{\sigma\sigma}$ as the projectors onto the respective irreducible subspace, than 
\begin{equation}
\mathcal{W}_{\mathcal{C}\otimes \mathcal{C}} (R_{\Lambda})_{ij}= \delta_{ij} \left\{\begin{array}{ccc}
1 &{\rm if}& P_i \in \{II \} \\
 \textrm{Tr} (R^{(\Lambda)} \mathbb{P}_{\mathbb{I}\sigma} )/3 &{\rm if}& P_i \in\{IX,IY,IZ \} \\
 \textrm{Tr} (R^{(\Lambda)} \mathbb{P}_{\sigma\mathbb{I}} )/3&{\rm if}& P_i \in\{XI,YI,ZI \} \\
\textrm{Tr} (R^{(\Lambda)} \mathbb{P}_{\sigma\sigma} )/9&& {\rm otherwise}. \\
\end{array} \right.
\end{equation}

Twirling over $\mathcal{C} \otimes I$ is quite a bit more complicated due the multiplicity of the distinct irreducible representation in the PTM.  In keeping with the conventions of the first section of this supplement let's label the irreducible representations of the PTM of $\mathcal{C} \otimes I$ as
\begin{equation}
\begin{split}
\sigma_{\mathbb{I},I} =& \{II\},\sigma_{\mathbb{I},X} = \{IX\},\sigma_{\mathbb{I},Y} = \{IY\},\sigma_{\mathbb{I},Z} = \{IZ\},\\
\sigma_{\sigma,I} =& \{XI,YI,ZI\}, \sigma_{\sigma,X} = \{XX,YX,ZX\},\\
\sigma_{\sigma,Y} =& \{XY,YY,ZY\}, \sigma_{\sigma,Z} = \{XZ,YZ,ZZ\}.
\end{split}
\end{equation}  
While we could derive twirling from Corollary 5, there is a more intuitive way to calculate this expression which we will show here.  

First we can express the two qubit $\mathcal{R}$ matrix in terms of the single qubit Pauli operators by doubling the indices,
\begin{equation}
(\mathcal{R}_\Lambda)_{ij,kl} =  \frac{1}{4}\mathrm{Tr}[P_i\otimes P_j \Lambda (P_k \otimes P_l)].
\end{equation}
and therefore the twirl is of the form
\begin{equation}
\mathcal{W}_{\mathcal{C}\otimes I}(\mathcal{R}_\Lambda)_{ij,kl} =\frac{1}{4|\mathcal{C}|}\sum_{U_m \in \mathcal{C}} \mathrm{Tr}[U_m P_i U_m^\dagger\otimes P_j \Lambda (U_m P_k U_m^\dagger \otimes P_l)].\label{eq:twirl_CI}
\end{equation}
From this expression we can show that if $i \neq k$ the matrix element is always zero by a symmetry arguement.  For any two Pauli operators $P_i \neq P_k$ (with $P_k \neq P_0$ for this example) there is always an element, $C$, of the Clifford group such that $C P_i C^\dagger = P_i$ and $C P_k C^\dagger = -P_k$.  This is due to the fact that the only constraint on the Clifford group is that it preserves commutation relations.  Substituting $U_m \rightarrow U_m C$ in Eq.~\eqref{eq:twirl_CI} yields $\mathcal{W}_{\mathcal{C}\otimes I}(\mathcal{R}_\Lambda)_{ij,kl} = -\mathcal{W}_{\mathcal{C}\otimes I}(\mathcal{R}_\Lambda)_{ij,kl} = 0$ for $i\neq j$.  For the non-zero components we note that the Clifford group preserves the ideneity and unfirmly redistributes $X$, $Y$ and $Z$ so that the full form of the twirl is
\begin{equation}
\mathcal{W}_{\mathcal{C}\otimes I} (R_{\Lambda})_{ij,kl}= \delta_{ik} \left\{\begin{array}{ccc}
(R_{\Lambda})_{0j,0l} &{\rm if}& i=0 \\
\sum_{n=1}^3 (R_{\Lambda})_{nj,nl} /3 && {\rm otherwise}. \\
\end{array} \right.
\end{equation}
The interpretation of this quantity is that for each configuration of the untwirled qubit we have a single qubit twirl.

\subsubsection{Unitary twirling}

Twirling over the entire unitary group yields exactly the same result as the Clifford group.